\documentclass[a4paper,11pt]{article}
\usepackage{jinstpub} % for details on the use of the package, please see the JINST-author-manual
\usepackage{lineno}
\usepackage{caption}
\usepackage{subcaption}
\usepackage{siunitx}
\usepackage{xspace}

\makeatletter
\newcommand{\oset}[3][0ex]{
  \mathrel{\mathop{#3}\limits^{
    \vbox to#1{\kern-2.5\ex@
    \hbox{$\scriptstyle#2$}\vss}}}}
\makeatother

\newcommand\overbar[1]{\oset[-0.2ex]{
   \textbf{--}}{#1}}

\newcommand{\numu}{\ensuremath{\nu_{\mu}}\xspace}
\newcommand{\numubar}{\ensuremath{\overbar{\nu}_{\mu}}\xspace}

\newcommand{\wbls}{WbLS\xspace}

\title{\boldmath Deployment of Water-based Liquid Scintillator in the Accelerator Neutrino Neutron Interaction Experiment}

 \collaboration{\includegraphics[height=17mm]{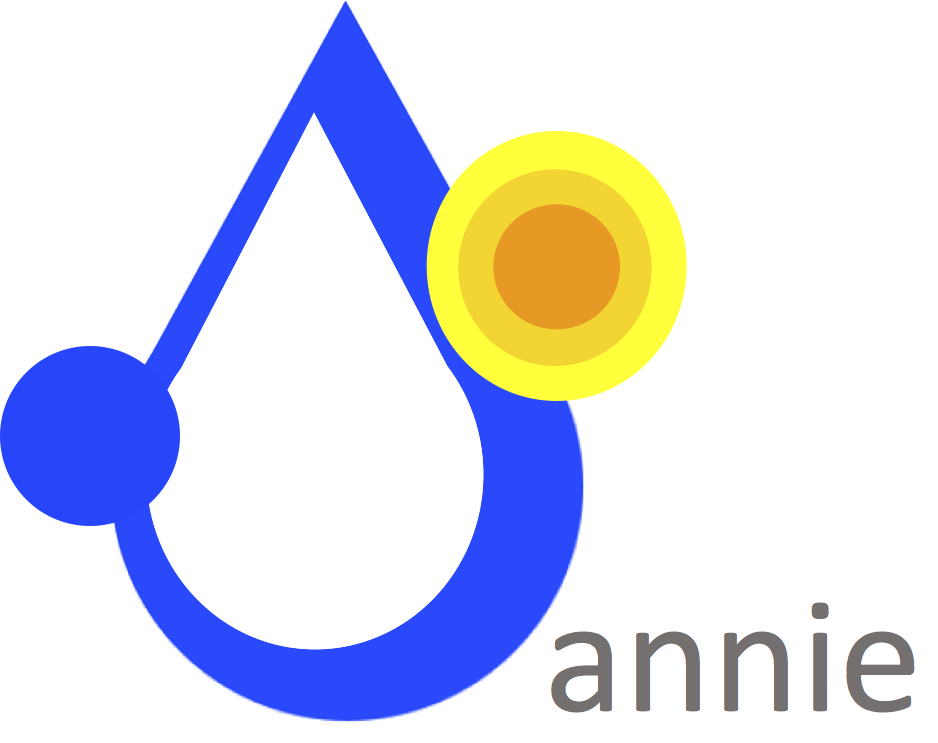}\\[6pt]
  The ANNIE collaboration}
\author[a]{M. Ascencio-Sosa,}
\author[b, c]{Z. Bagdasarian,}
\author[d]{J. F. Beacom,}
\author[e]{M. Bergevin,}
\author[f]{M. Breisch,}
\author[g]{G. Caceres Vera,}
\author[e]{S. Dazeley,}
\author[a]{S. Doran,}
\author[h]{E. Drakopoulou,}
\author[a]{S. Edayath,}
\author[i]{R. Edwards ,}
\author[j]{J. Eisch,}
\author[a]{Y. Feng,}
\author[j]{V. Fischer,}
\author[k]{R. Foster,}
\author[j]{S. Gardiner,}
\author[l]{S. Gokhale,}
\author[g]{P. Hackspacher,}
\author[m]{C. Hagner,}
\author[g]{J. He,}
\author[f]{B. Kaiser,}
\author[a]{F. Krennrich,}
\author[f]{T. Lachenmaier,}
\author[n]{F. Lemmons,}
\author[o]{D. Maksimovic,}
\author[k]{M. Malek,}
\author[o, *]{J. Martyn,\note[*]{Corresponding author.}}
\author[p]{A. Mastbaum,}
\author[j]{C. McGivern,}
\author[p]{J. Minock,}
\author[o]{M. Nieslony,}
\author[i]{M. O'Flaherty,}
\author[b,c]{G. D. Orebi Gann,}
\author[e]{T. Pershing,}
\author[b,c]{L. Pickard,}
\author[q]{N. Poonthottathil,}
\author[l]{C. Reyes,}
\author[i]{B. Richards,}
\author[l]{R. Rosero,}
\author[r]{M. C. Sanchez,}
\author[o]{D. T. Schmid,}
\author[s]{M. Smy,}
\author[m]{M. Stender,}
\author[r]{A. Sutton,}
\author[g]{R. Svoboda,}
\author[t, u]{E. Tiras,}
\author[s]{M. Vagins,}
\author[a]{V. Veeraraghavan,}
\author[n]{J. Wang,}
\author[a]{A. Weinstein,}
\author[a]{M. Wetstein,}
\author[o]{M. Wurm,}
\author[l]{M. Yeh,}
\author[g]{T. Zhang}

\affiliation[a]{Iowa State University, Department of Physics and Astronomy, Ames, IA 50011 U.S.A.}
\affiliation[b]{University of California, Berkeley, Physics Department, Berkeley, CA 94720 U.S.A.}
\affiliation[c]{Lawrence Berkeley National Laboratory, Nuclear Science Division, Berkeley, CA 94720 U.S.A.}
\affiliation[d]{The Ohio State University, Department of Physics, Columbus, OH 43210 U.S.A.}
\affiliation[e]{Lawrence Livermore National Laboratory, Livermore, CA 94550 U.S.A.}
\affiliation[f]{Eberhard Karls Universit\"at, Kepler Center for Astro and Particle Physics, T\"ubingen 72076, Germany}
\affiliation[g]{University of California at Davis, Department of Physics and Astronomy, Davis, CA 95616, U.S.A.}
\affiliation[t]{Erciyes University, Department of Physics, Kayseri, 38030, T\"urkiye}
\affiliation[h]{N.C.S.R. "Demokritos", Institute of Nuclear and Particle Physics, Agia Paraskevi 15341, Greece}
\affiliation[i]{University of Warwick, Department of Physics, Coventry CV4 7AL U.K.}
\affiliation[j]{Fermi National Accelerator Laboratory, Batavia, IL 60510, U.S.A.}
\affiliation[k]{University of Sheffield, Department of Physics and Astronomy, Sheffield, S10 2TN, U.K.}
\affiliation[o]{Johannes Gutenberg Universit\"at, Institut f\"ur Physik, Mainz 55128, Germany}
\affiliation[l]{Brookhaven National Laboratory, Upton, NY 11973, U.S.A.}
\affiliation[m]{Universit\"at Hamburg, Institut f\"ur Experimentalphysik, Hamburg 22761, Germany}
\affiliation[n]{South Dakota School of Mines and Technology, Physics Department,  Rapid City SD, 57701 U.S.A.}
\affiliation[p]{Rutgers University, Department of Physics and Astronomy, Piscataway, NJ 08854 U.S.A.}
\affiliation[q]{Indian Institute of Technology Kanpur, Department of Physics, Kanpur 208016, India}
\affiliation[r]{Florida State University, Department of Physics, Tallahassee, FL 32306 U.S.A.}
\affiliation[s]{University of California at Irvine, Department of Physics and Astronomy, Irvine CA, 92697 U.S.A.}
\affiliation[u]{University of Iowa, Department of Physics and Astronomy, Iowa City, IA 52242 U.S.A.}

\emailAdd{jomartyn@uni-mainz.de}

\abstract{
The Accelerator Neutrino Neutron Interaction Experiment (ANNIE) is a 26-ton water Cherenkov neutrino detector installed on the Booster Neutrino Beam (BNB) at Fermilab.
Its main physics goals are to perform a measurement of the 
neutron yield from neutrino-nucleus interactions, as well as a measurement of the charged-current cross section of muon neutrinos.
An equally important focus is the research and development of new detector technologies and target media.
Specifically, water-based liquid scintillator (\wbls) is of interest as a novel detector medium, as it allows for the simultaneous detection of Cherenkov light and scintillation. 
This paper presents the deployment of a 366\,L \wbls vessel in ANNIE in March 2023 and the subsequent detection of both Cherenkov light and scintillation from the \wbls.
This proof-of-concept allows for the future development of reconstruction and particle identification algorithms in ANNIE, as well as dedicated analyses within the \wbls volume, such as the search for neutral-current events and the hadronic scintillation component.
}


\begin{document}
\maketitle
\flushbottom

\section{Introduction}

The Accelerator Neutrino Neutron Interaction Experiment (ANNIE)~\cite{annie_2017, annie_2020} is a 26-ton gadolinium loaded water Cherenkov detector deployed on the Booster Neutrino Beam (BNB) at Fermilab.
The physics goals of ANNIE focus on, but are not limited to, the measurement of the final-state neutron multiplicity produced by \numu interactions with nuclei, as well as the measurement of the charged-current cross section of \numu.
Additionally, ANNIE is a test platform for novel detection concepts and therefore has a strong focus on different detector research and development tasks. 
These include the deployment of Large Area Picosecond Photodetectors (LAPPDs)~\cite{LAPPD_2015, LAPPD_2020} as a new photodetector technology, as well as Water-based Liquid Scintillator (\wbls)~\cite{WbLS_2020, WbLS_2011} as a novel detector medium, which is the focus of this paper.

Given the success of monolithic, large scale water Cherenkov and liquid scintillator detectors in the context of neutrino physics, there is currently an ongoing interest in the concept of hybrid event detection.
The aim is to combine the advantages of water Cherenkov detectors --- large optical transparency, sensitivity to the particle direction, as well as particle identification through the Cherenkov ring topology --- with the low-threshold, calorimetric information of liquid scintillator detectors.
The latter would provide a better energy resolution, additional particle identification capabilities and allow for the detection of neutral particles or particles with energies below the Cherenkov threshold.
To achieve such a hybrid event detection it is necessary to differentiate between the Cherenkov and scintillation components of the events.

A number of different approaches have been developed for this purpose in recent years, such as spectral photon sorting~\cite{bandpass_2019, dichroicon_2020}, time separation through fast photodetectors~\cite{Aberle_2014, fast_timing_2017, fast_timing_2019} or the development of novel target materials, such as slow scintillators~\cite{slow_scint_2019, slow_scint_2022} and \wbls~\cite{WbLS_2020, WbLS_2011}.
It is also possible to combine these separation approaches, such as the use of fast LAPPDs together with a \wbls medium~\cite{wbls_lappd_2022}.
\wbls is an admixture of a low percentage of liquid scintillator in water, featuring high transparency and which is an enabling medium for hybrid neutrino event detection.
Compared to pure liquid scintillator, it has a reduced scintillation yield, which allows for easier identification of the Cherenkov light.
The \wbls hence provides a larger total number of photons compared to Cherenkov light from pure water.
Conceptually similar ideas have already been used successfully in LSND~\cite{LSND_1993} and in MiniBooNE~\cite{MiniBooNE_reco_2009}, using a diluted scintillator and mineral oil as detection media, respectively.

ANNIE is part of a wider experimental effort for the technical demonstration of the hybrid optical detection concept.
This includes bench-top scale experiments such as CHESS~\cite{chess_2017, WbLS_2020} and FlatDot~\cite{fast_timing_2019} and also other mid-scale detectors setups: Eos~\cite{eos_2023}, the 1-ton~\cite{bnl_1ton_prep} and 30-ton purification demonstrator at Brookhaven, and NuDot~\cite{FSNN_whitepaper_2023}.
Furthermore \wbls is explored as a detection medium for the proposed THEIA detector~\cite{theia_2020}.
Several recent community-planning exercises have been highlighting the importance of the developments summarized above~\cite{FSNN_whitepaper_2023, snowmass_2022_a, snowmass_2022_b}.

This paper provides the first results on the deployment of a small \wbls vessel in ANNIE and it presents the first detection of accelerator neutrinos in a \wbls medium. 
Section~\ref{sec:annie_detector} gives a short overview of the ANNIE detector, while section~\ref{sec:SANDI_deployment} explains the mechanical structure, filling and deployment of the \wbls vessel inside the ANNIE water tank. 
Section~\ref{sec:wbls_description} provides a description of the \wbls deployed in ANNIE and shows its long term stability.
Section~\ref{sec:SANDI_detection} shows a data-on-data comparison of events with and without the \wbls scintillation.
This provides a proof for the successful deployment of \wbls in ANNIE, corresponding to the detection of both Cherenkov and scintillation light associated with the accelerator neutrino events.
Conclusions are given in section~\ref{sec:conclusion}.

\section{The ANNIE detector}
\label{sec:annie_detector}

\begin{figure}[htb!]
    \centering
    \includegraphics[width=0.75\textwidth]{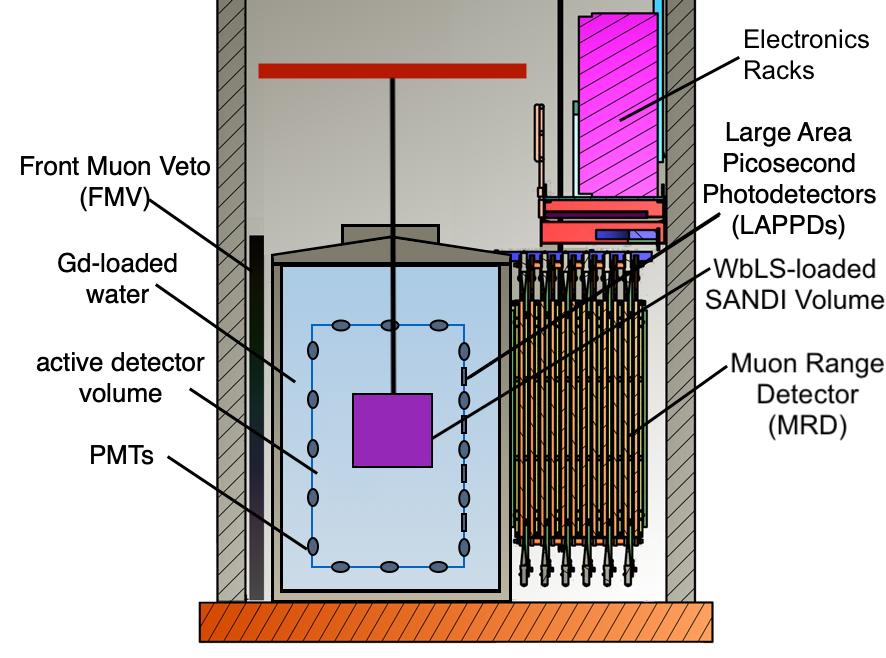}
    \caption{Detector schematic with the \wbls-filled vessel, called SANDI (Scintillator for ANNIE Neutrino Detection Improvement).}
    \label{fig:anniedetector}
\end{figure}

ANNIE consists of three separate main detector elements: (1) the front muon veto (FMV), (2) the main water Cherenkov detector, and (3) the muon range detector (MRD), as shown schematically in Figure~\ref{fig:anniedetector}. The ANNIE hall~\cite{annie_2017} consists of three levels (top, middle, and lower). The top is at ground level and is the entrance to the hall with the middle and lower levels below ground level. The middle level houses the electronics racks and has access to the lid of the ANNIE tank.

The FMV, located upstream of the main detector volume, consists of two layers of scintillating paddles. Each paddle is read out at one end by a 2-inch PMT, coupled to a glass light guide. The primary purpose of the FMV is to reject events originating from an interaction in the dirt upstream of ANNIE. Only the existence of an incoming particle is relevant and not the exact position, therefore both layers are oriented horizontally. The two layers are read out on opposite ends to mitigate the impact of attenuation losses from a particle passing through the far end of a paddle. The FMV was found to have a muon tagging efficiency of $\left(95.6 \pm 1.6 \right)\%$~\cite{nieslony_2022}.
This efficiency was determined through the use of the MRD muon track reconstruction,
by back-propagating found muon tracks to the position of the FMV and checking for a coincidence with a FMV paddle hit.

The MRD is placed downstream of the main detector volume, and like the FMV, consists of layers of \SI{6}{\milli\meter} thick scintillator paddles read out by 2-inch PMTs, arranged in alternating horizontal and vertical orientations sandwiched by layers of \SI{5}{\milli\meter} thick steel absorbers. This arrangement provides the capability to determine the muon direction through the MRD. In total there are 6 horizontal and 5 vertical layers of scintillator paddles with 11 layers of steel. The PMTs are a combination of EMI 9954KB and RCA 6342A. The RCA PMTs have about one-tenth the gain of the EMI PMTs and are therefore amplified by a factor a 10 in the NIM readout logic.

The primary detector of ANNIE is a 26-ton, cylindrical, highly pure water volume loaded with 0.1\% of gadolinium by weight~\cite{gadolinum_2004}. The tank measures 13-ft tall by 10-ft in diameter, made of 7-gauge (4.5\,mm) steel. The interior of the tank is instrumented with 132 PMTs for a 10\% photo cathode coverage.
The ANNIE tank has a removable top hatch measuring about \SI{1}{\meter} in diameter, which supports four PMTs. Additionally, the tank lid has several small ports through which calibration sources can be lowered into the water.

%The primary detector of ANNIE is a 26-ton, cylindrical, highly pure water volume loaded with 0.2\% of gadolinium sulfate by weight. The tank measures 13-ft tall by 10-ft in diameter, made of 7-gauge (4.5 mm) steel. The interior of the tank is instrumented with 132 PMTs for 10\% photo cathode coverage. Five Large Area Picosecond Photodetectors (LAPPDs), provided by Incom Inc., will supplement the PMTs to provide improved timing and spatial resolution. LAPPDs are \SI{20}{\centi\meter}$~\times~$\SI{20}{\centi\meter} photodetectors that utilize micro-channel plates (MCPs) to achieve sub-centimeter spatial resolution and a timing resolution down to the order of tens of picoseconds \cite{LAPPD_2020}. The ANNIE tank has a removable top hatch measuring about \SI{1}{\meter} in diameter, which supports four PMTs. Additionally, the tank lid has several small ports where calibration sources can be lowered into the water and slots through which LAPPDs can be deployed.

ANNIE is placed \SI{100}{\meter} downstream of the BNB target. The BNB is produced by impinging \SI{8}{\giga\eV} protons onto a beryllium target. The resulting hadrons are focused by a toroidal magnetic horn that is typically pulsed to focus positive hadrons while defocusing negative ones. These hadrons, primarily pions, then decay to produce neutrinos. The associated charged leptons are stopped by an absorber \SI{50}{\meter} downstream of the target. The BNB produces a flux consisting of 93.6\% \numu and 5.9\% \numubar with a spectrum peaked around \SI{700}{\mega\eV} \cite{MiniBooNE_2009}, and operates at a maximum average rate of \SI{5}{\hertz}, delivering about $4 \times 10^{12}$ protons on target (POT) over a \SI{1.6}{\micro\second} spill. 

\section{Deployment of \wbls in ANNIE}
\label{sec:SANDI_deployment}
\wbls was deployed in ANNIE from March to May of 2023 in a program dubbed SANDI (Scintillator for ANNIE Neutrino Detection Improvement). The SANDI vessel is a cylinder made of \SI{2.54}{\centi\meter} thick acrylic with an interior measuring \SI{90}{\centi\meter} in height and \SI{72}{\centi\meter} in diameter giving it a \SI{366}{\liter} capacity. The vessel is held by a stainless steel frame, which serves as a lifting structure. Electro-polished SAE 304 stainless steel is used to ensure compatibility with the gadolinium-loaded water. With the stainless steel structure, the vessel weighs $\sim$\SI{163}{\kilo\gram} when empty. After fabrication, the vessel was leak tested by pressurizing it to \SI{2}{psi} for 24 hours, during which time the vessel maintained this pressure. 

\begin{figure}[h!]
	\begin{center}
		\includegraphics[width=.65 \linewidth]{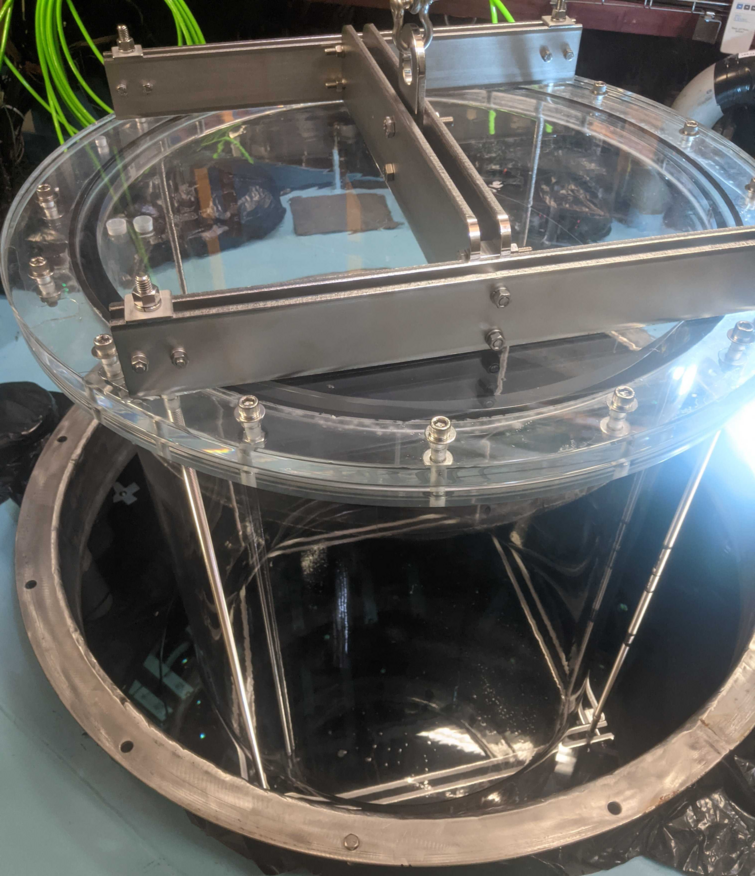} 
	\end{center}
	\caption{The SANDI vessel suspended from the deployment structure.}
		\label{fig:sandi}
\end{figure}

To deploy SANDI, a steel support structure was designed, fabricated, and installed by the Fermilab engineering staff.
The structure consists of two 4-inch square steel support tubes with 1/4-inch thick walls. A W8$\times$13 steel I-beam spans the columns to support the SANDI vessel. Connected to the I-beam is an electrical winch with a Teflon-wrapped stainless steel cable. The winch is on rollers and was used to lift the vessel, maneuver it into position, and lower it into the water. Figure~\ref{fig:sandi} shows the vessel, hanging from the deployment structure. 

Filling the SANDI vessel took two days to complete. Two days prior to filling, the SANDI vessel was flushed with nitrogen to reduce the potential for biological contamination. 
To account for fluid displacement, gadolinium-loaded water was removed from the tank, stored in two drums lined with polyethylene, and later reused after extracting SANDI. The top hatch of the ANNIE tank, along with its four PMTs, was removed and stored on the tank lid. 
The empty SANDI vessel was then lowered into the center of the tank, just to the point where it became buoyant. As the vessel was filled, it was lowered in steps to maintain access while limiting the additional force on the support structure. 

Due to the small size of the ANNIE hall, the \wbls barrels were held outside of the ANNIE hall under a tent.
The \wbls was pumped from the barrels through a 0.2-micron filter and into \SI{10}{\liter} containers. The containers were then carried down to the lower level where they were emptied through a funnel connected to the inlet of the SANDI vessel. In addition to the inlet port, the SANDI vessel was also fitted with a Teflon overflow tube. When the vessel was nearly full, the inlet was capped off with a stainless steel fitting, and the overflow tube was filled with about \SI{3}{\meter} of \wbls to remove the last air pocket. Once the vessel was lowered to its final location, a temporary PVC lid with a cutout for the suspension cable and overflow tube was fitted to the ANNIE tank. The overflow tube and temporary lid were both covered with opaque plastic. 

During the deployment period, water samples were extracted from the ANNIE tank for observation. The samples were analyzed with a UV-vis spectrophotometer to monitor the absorption spectrum and check for \wbls contamination. The resin filters of the ANNIE water system were expected to capture \wbls from any small leaks. Throughout the duration of the deployment, the UV-vis monitoring did not show any change in the absorption spectrum that would indicate \wbls contamination of the main water tank. 

The extraction of the SANDI vessel proceeded much the same way as the deployment. However, in this case we pumped directly from the vessel on the middle level of the ANNIE hall up to the \wbls storage barrels on the top level. Samples of the extracted \wbls were taken back to Brookhaven National Laboratory (BNL) for analysis.

\section{Production and long term stability of the deployed \wbls}
\label{sec:wbls_description}
The organic component of the \wbls used in SANDI is a mix of modified polyethylene glycol-based surfactants and a diisopropylnaphthalene (DIN) base liquid scintillator loaded with a fluor of 2,5-diphenyloxazole (PPO). 
All raw components were purified using vacuum distillation and an exchange column before synthesis. To fill the SANDI tank, approximately 500\,L of \wbls were produced using an in-situ mixing technique~\cite{bnl_1ton_prep} with a double-jacketed 90\,L Chemglass reaction vessel at BNL.
This \wbls formula was designed to be compatible with gadolinium loading, similar to the \wbls that was deployed in CHESS for precision light-yield and time profile measurements~\cite{chess_2023_prep}, and at the BNL 1-ton test-bed for optical transparency, stability, and larger-scale testing~\cite{bnl_1ton_prep}. 
This \wbls formulation differs from the one studied in~\cite{WbLS_2011, WbLS_2020}, in order to provide compatibility with gadolinium loading, and therefore the optical parameters are expected to differ.

The reaction vessel for the production of the \wbls has several injection ports made of polytetrafluoroethylene (PTFE) to prevent any chemical complication of adding organic materials. All the tubing, filtration system, liners, and the mixing system were pre-cleaned with ethanol (Ethyl alcohol, 190 proof), rinsed with ultra pure water (resistivity 18.2 M$\Omega\cdot\text{cm}$) and dried with nitrogen gas. Afterward, it was sealed in an inert environment until use. 
At different stages of the synthesis, 1\% (mass) of purified organic materials were introduced through different ports into the reactor where pure water (99\% by mass) was pre-filled and mixed at 100\,rpm for four hours. Finally, the \wbls was filtered through a 0.2-micron PTFE membrane using a 316-stainless steel filtration housing and stored in 55-gal lined drums for shipment.

Each drum was outfitted with a dual polyethylene liner, each measured 5-micron in thickness, to serve as a reservoir for liquids. The maximum storage capacity of each drum was set at 180 L, equivalent to two 90 L batches from the vessel system, to facilitate handling and to prevent overflow. Nitrogen cover gas was introduced to the liquid to ensure its stability during transport. Three \wbls drums ($\approx$450\,L) were shipped to and placed in a temperature-controlled warehouse at Fermilab in November 2022.

\begin{figure}[htb!]
    \centering
    \begin{subfigure}[b]{0.49\textwidth}
        \centering
        \includegraphics[width=\textwidth]{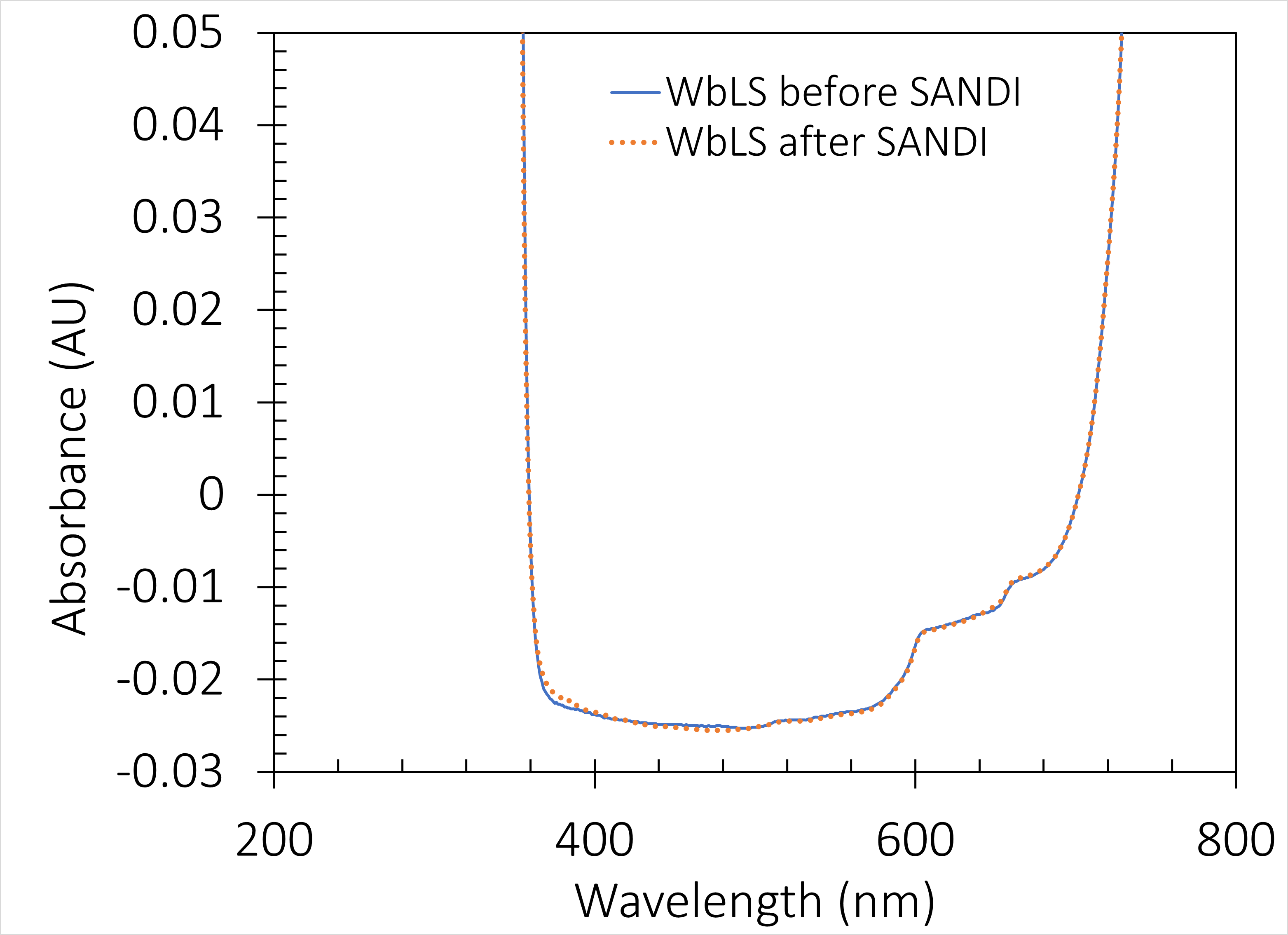}
        \caption{}
        \label{fig:WbLS_optical}
    \end{subfigure}
    \begin{subfigure}[b]{0.49\textwidth}
        \centering
        \includegraphics[width=\textwidth]{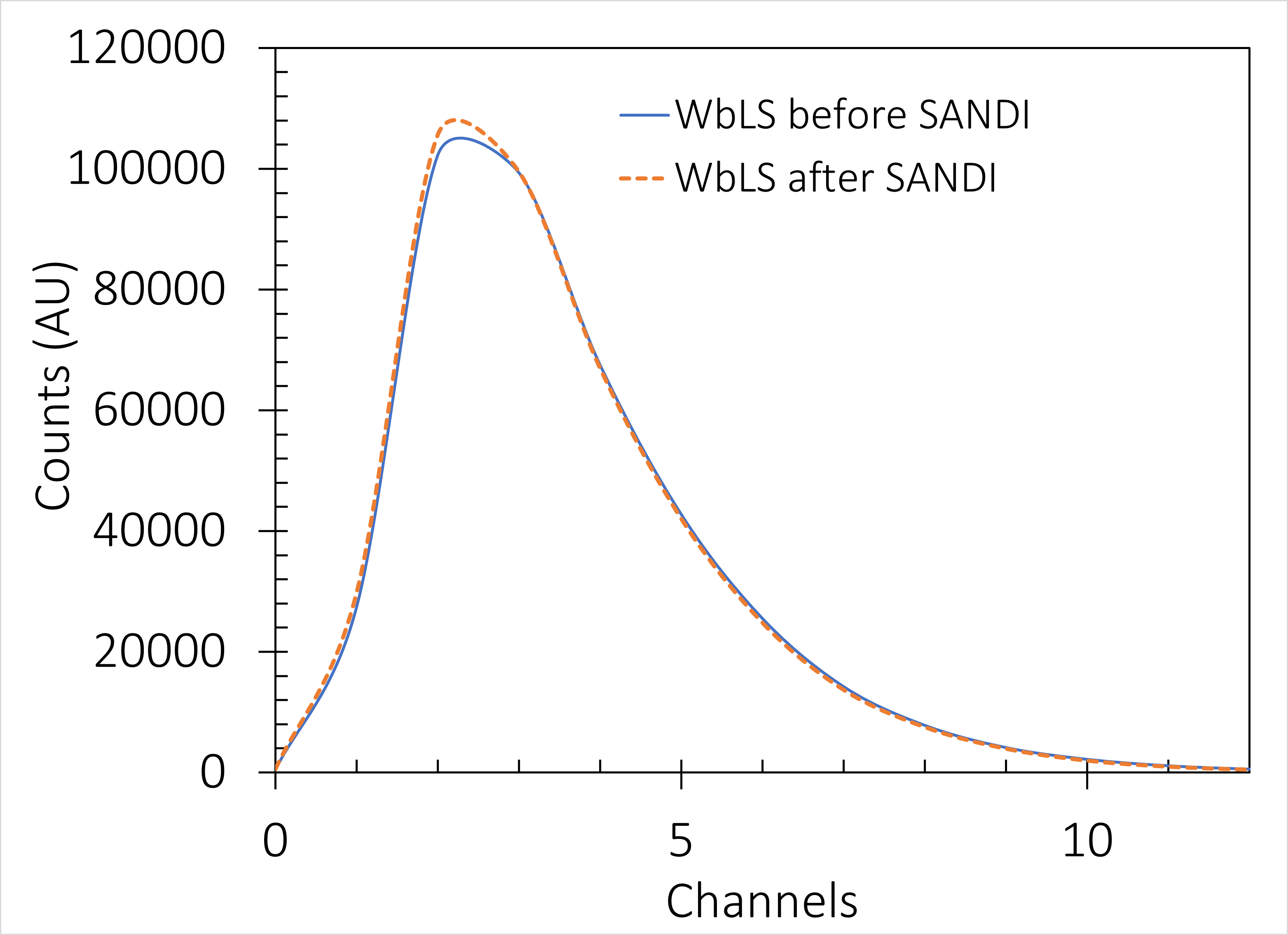}
        \caption{}
        \label{fig:WbLS_LY}
    \end{subfigure}
    \caption{(a) The \wbls optical transparency before and after the SANDI deployment. 
    (b) The \wbls light-yield before and after the SANDI deployment, given by the Compton spectra induced by $^{137}$Cs.}
    \label{fig:wbls_optical_ly}
\end{figure}

The performance of the \wbls used in SANDI was assessed using a Shimadzu UV-vis spectrophotometer with a 10\,cm quartz cell for optical transmission and a Beckman LS6500 coincidence counter using a $^{137}$Cs source for a light-yield measurement at BNL. Figure~\ref{fig:WbLS_optical} shows the optical stability before and after the SANDI deployment. The variation in optical transmission is within the statistic uncertainty (0.0003) of the two baselines.
Figure~\ref{fig:WbLS_LY} shows the Compton spectra induced from $^{137}$Cs.
This represents the light-yield stability of the \wbls before and after its deployment, as the spectra are the same within the measurement uncertainty.
Together, the UV and light-yield data represent the overall \wbls stability through different processes, including shipping, storage, deployment, and draining, over a nine-month period since production.

\section{Detection of \wbls scintillation in ANNIE}
\label{sec:SANDI_detection}
This section describes the detection of the \wbls scintillation light in ANNIE from the BNB muon-neutrinos.
Two independent analyses are performed to estimate the effect of the \wbls-filled SANDI vessel on the number of detected photoelectrons (p.e. or charge) in ANNIE, relative to the Cherenkov light of pure water.
This is done through a data-on-data comparison of throughgoing muon events and Michel electrons, which is meant to be a principal proof for the \wbls feasibility in a neutrino beam.
The two data sets used in this section are the pre-SANDI data (without \wbls), running from  December 2022 to  March 2023, and the SANDI data (with \wbls), from  March 2023 to  May 2023.

\subsection{Throughgoing muon analysis}
Muons can be produced by the interaction of the beam $\nu_\mu$ with nuclei.
If such a muon is produced outside of SANDI, it can cross the ANNIE detector from front to back and is called a throughoing muon.
These events are well suited to estimate the impact of the \wbls-filled SANDI vessel on the number of detected photoelectrons and the additional contribution of the scintillation.
The reason for this is that these muons are minimally ionizing particles which produce a relatively well defined photoelectron spectrum, depending on the entrance angle and total track length in ANNIE, smeared by the detector response.

The principal event selection of neutrino-associated muon events is performed in a data acquisition window of $\Delta t = 2 \,\mu\text{s}$ of every incoming beam trigger. 
First, a coincidence is required between the water tank cluster and the MRD cluster within $100$\,ns to select events with a muon in the final state.
A water tank cluster is defined as a minimum of five tank PMT hits within a pre-defined time window of 50\,ns and a MRD cluster is defined as at least four MRD PMT hits in a time window of 30\,ns.
The MRD coincidence guarantees the selection of a muon event and the MRD is sensitive to the muon track direction.
Events are required to have exactly one reconstructed muon track to simplify the comparison analysis.
The MRD track is extrapolated back through the ANNIE water tank and events are selected if the muon track intersects the SANDI volume.
This is expected to provide a large number of selected events which can produce scintillation light, as they pass the SANDI volume. 
A similar cut on the muon tracks was performed for the ANNIE data without the SANDI vessel, where muon tracks were required to pass through the space where SANDI would be.
It has to be noted here that the MRD track angle reconstruction has an estimated uncertainty of 0.17\,rad ($9.7\circ$) and an uncertainty on the MRD entry point of about $12$\,cm in both the X and Y coordinates~\cite{nieslony_2022}.
This means that not all selected events have a true intersection with the SANDI volume.
Additionally, a lower limit on the detected number of photoelectrons in the water tank of $200$\,p.e.\ is selected to minimize the impact of noise.

These data sets are further split in two, depending on whether the FMV trigger is coincident with the PMT tank trigger or not.
When such a coincidence occurs between the FMV, the water tank and the MRD, the corresponding event can be identified as a throughgoing muon.
This event selection is well suited for an estimation of the impact of the additional scintillation, as the throughgoing muons provide a relatively simple event topology.
Muons that go through the entire ANNIE detector provide an energy distribution that is more narrow than muons that get produced inside the tank, because the spread in track length is smaller for the throughgoing muons.
If the FMV is not triggered in coincidence with the water tank and the MRD the event is considered a neutrino candidate, i.e. the neutrino interacts inside the water tank and the associated muon is detected by the MRD. 
These selection cuts provide four data sets with 1771 pre-SANDI neutrino candidate events, 1181 SANDI neutrino candidate events, 983 pre-SANDI throughgoing muon events, and 615 SANDI throughgoing muon events.

\begin{figure}[htb!]
    \centering
    \begin{subfigure}[b]{0.49\textwidth}
        \centering
        \includegraphics[width=\textwidth]{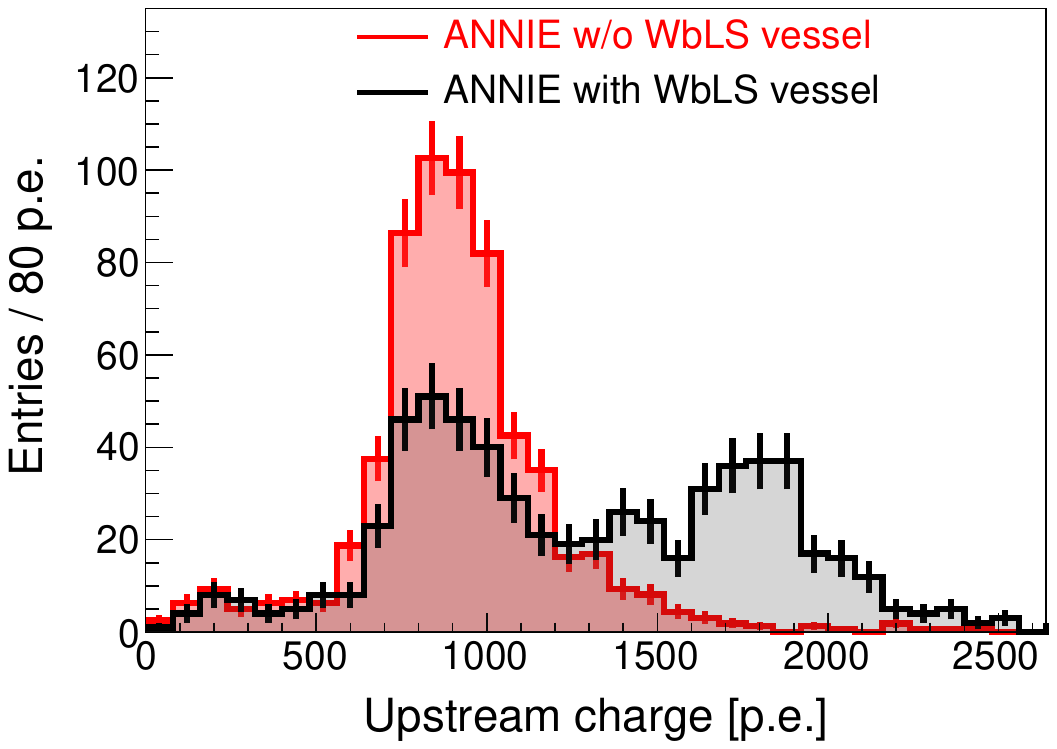}
        \caption{}
        \label{fig:comparison_up}
    \end{subfigure}
    \begin{subfigure}[b]{0.49\textwidth}
        \centering
        \includegraphics[width=\textwidth]{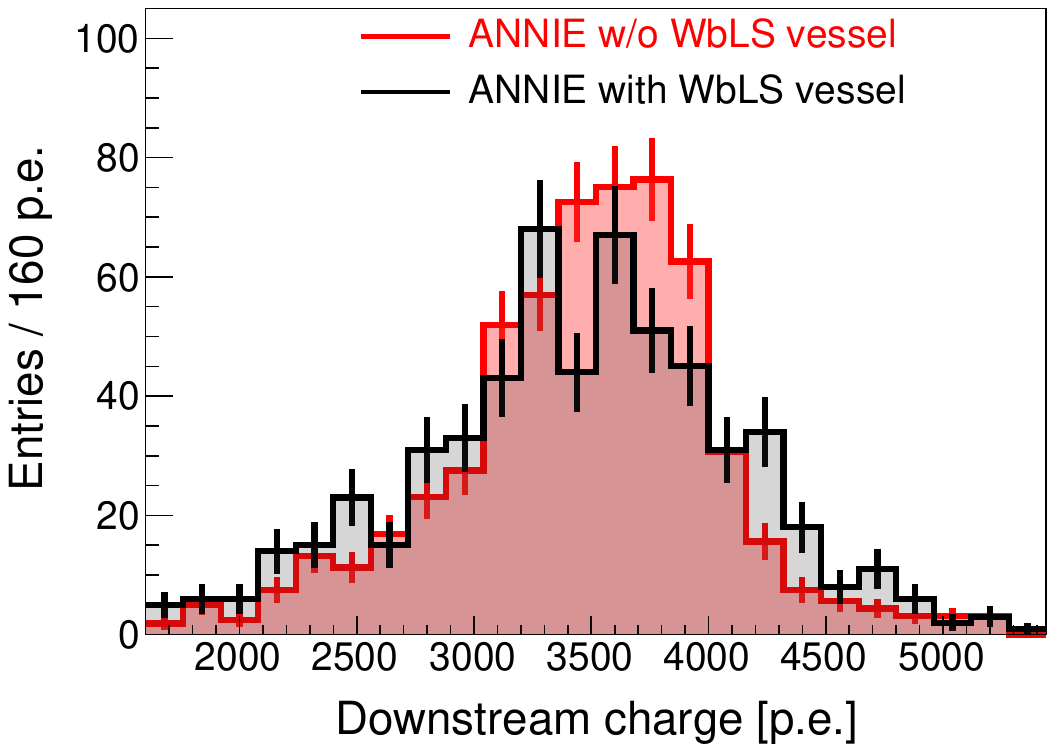}
        \caption{}
        \label{fig:comparison_down}
    \end{subfigure}
    \caption{Distribution of the charge for the selected throughgoing muon events in units of the number of detected photoelectrons for (a) upstream PMTs and (b) downstream PMTs. The events are selected if they have an intersection of the reconstructed muon track of ANNIE with the \wbls vessel volume. The black distributions show the data of ANNIE with the \wbls vessel and the red distributions show the data of ANNIE without the \wbls vessel. The latter distribution is normalized to the statistics of the black distribution of the ANNIE with \wbls events.}
    \label{fig:throughgoing_comparison}
\end{figure}

The water tank PMTs are divided into two categories: The upstream PMTs with a Z-position $ < 0$\,m and the downstream PMTs with a Z-position $ > 0$\,m, where the Z-axis is defined by the neutrino beam direction and the origin of the coordinates is the center of the water tank.
For Cherenkov light it is expected that the downstream PMTs should detect a relatively large number of photoelectrons, as it is emitted in a forward cone along the muon direction. 
The upstream PMTs are not able to see direct Cherenkov light, but they can detect a small fraction of the downstream photoelectrons coming from reflections off the ANNIE structures and downstream PMT glass.
The scintillation from the \wbls vessel on the other hand is emitted isotropically. 
It is therefore expected that both the upstream and downstream PMTs should detect a substantial amount of direct scintillation light, while seeing a very different amount of direct or indirect Cherenkov light.
This effect can be seen for the throughgoing muon data sets in  Figure~\ref{fig:throughgoing_comparison}.

Figure~\ref{fig:comparison_up} shows the measured distribution for the total number of detected photoelectrons (charge) of all upstream PMTs for the SANDI data (ANNIE with the \wbls vessel) in black and for the pre-SANDI data (without the \wbls vessel) in red.
The latter distribution is normalized in the plot to the statistic of the SANDI data for better comparison.
It can be seen that the pre-SANDI data has a charge distribution with a single peak around $900$\,p.e., which corresponds to the reflected Cherenkov light of muons traversing the entire tank.
The upstream charge distribution of the SANDI data shows two peaks at around $900$\,p.e. and $1700$\,p.e.
These two peaks originate from the non-perfect MRD track reconstruction, which gives rise to two populations of events: (1) Those events for which the reconstructed muon track has an intersection with the \wbls vessel, while the true muon track does not intersect with the \wbls vessel.
(2) Those events for which both the reconstructed and true muon track do have an intersection with the \wbls vessel.
Here, a more sophisticated track direction and interaction vertex reconstruction could provide a better event selection, by using the charge and time information of the tank photodetectors. 
Such a technique is under development.
The peak with a lower charge corresponds to events of the first type, which did not produce scintillation light, while the peak at the larger charge includes both the reflected Cherenkov light and the direct scintillation light.
Therefore, Figure~\ref{fig:comparison_up} provides a qualitative demonstration of the detection of scintillation from the \wbls.
It is interesting to note that the charge distribution of the muons that miss the \wbls has a peak position that is comparable to that of the distribution for the pre-SANDI data events.
This likely indicates that the reflection of Cherenkov light from the SANDI vessel structure is relatively small.
If the SANDI vessel structure would have reflected a large fraction of the Cherenkov photons, then the average upstream charge for the muon events that miss the \wbls would have been shifted noticeably, compared to the pre-SANDI data muon events.
This observation has been qualitatively confirmed by a simple simulation of the detector geometry.
It indicates that the reflections off the SANDI vessel structure can only provide a few-percent level effect relative to the detected number of downstream photons.

Figure~\ref{fig:comparison_down} shows the downstream charge distribution for ANNIE without the \wbls vessel in red and for ANNIE with the \wbls vessel in black.
Here, both distributions have a peak around $3400$\,p.e., where the presence of the \wbls vessel provides a distribution that is somewhat broadened compared to the distribution for pure Cherenkov light. 
This can be seen by comparing the root mean square of the two histograms, which is around 590\,p.e. for the pre-SANDI data and around 710\,p.e. for the SANDI data.
As expected, the number of detected photoelectrons for the downstream PMTs is larger than for the upstream PMTs, due to the directionality of the Cherenkov light from the selected muons.
The effect of the broadening for ANNIE with the \wbls vessel can be explained by the two event populations described previously.
Those muon tracks that are mis-reconstructed, such that the true muon track does not intersect the \wbls vessel, do not produce scintillation light.
At the same time, their Cherenkov photons can still traverse the \wbls, which has a lower transparency than the pure water.
Therefore, these events are shifted to a lower average number of downstream p.e., compared to the events for ANNIE without the \wbls.
Those events for which the muon tracks do have an intersection with the \wbls vessel produce both scintillation and Cherenkov light, shifting the downstream p.e.~distribution to higher values compared to ANNIE without the \wbls.
This is assumed to provide a broader downstream charge distribution for the SANDI data, as the contribution of the scintillation is not large enough compared to the direct Cherenkov hits to produce two distinctly separate peaks.

\begin{figure}[htb!]
    \centering
    \begin{subfigure}[b]{0.49\textwidth}
        \centering
        \includegraphics[width=\textwidth]{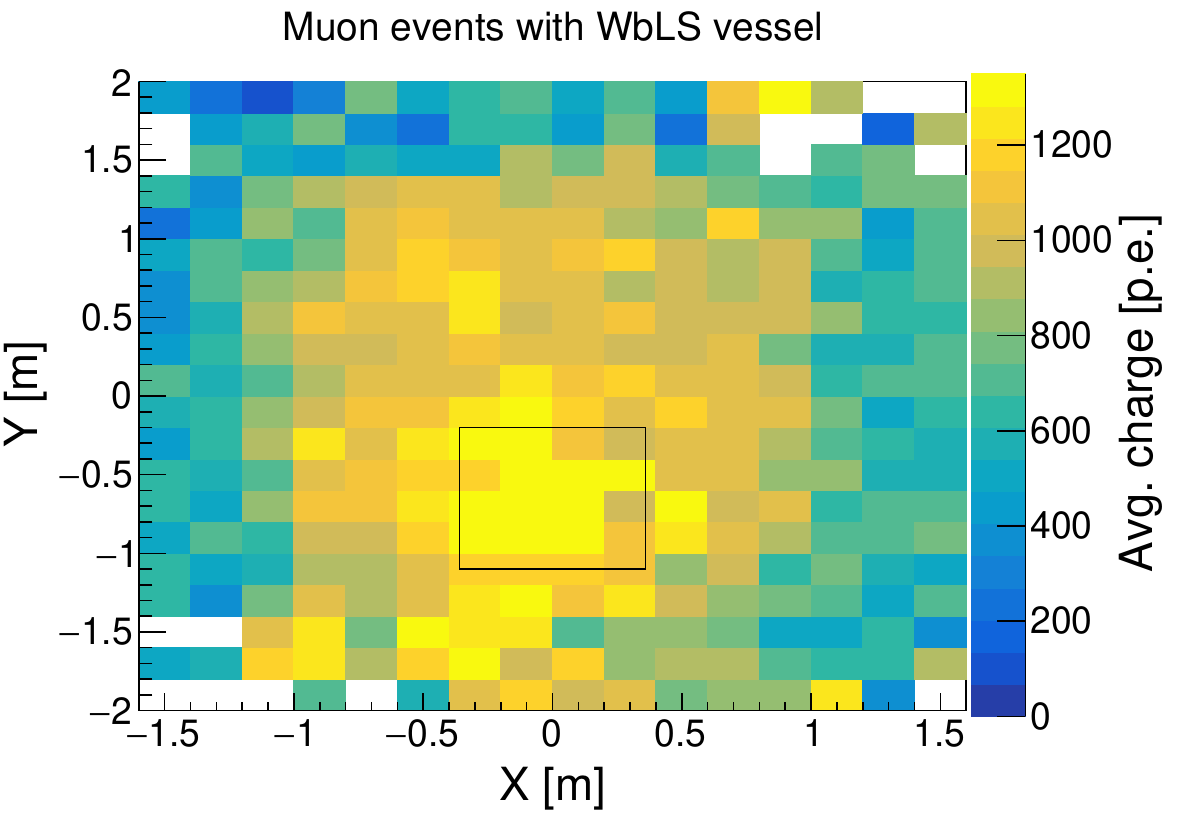}
        \caption{}
        \label{fig:muon_pos_vs_charge_sandi}
    \end{subfigure}
    \begin{subfigure}[b]{0.49\textwidth}
        \centering
        \includegraphics[width=\textwidth]{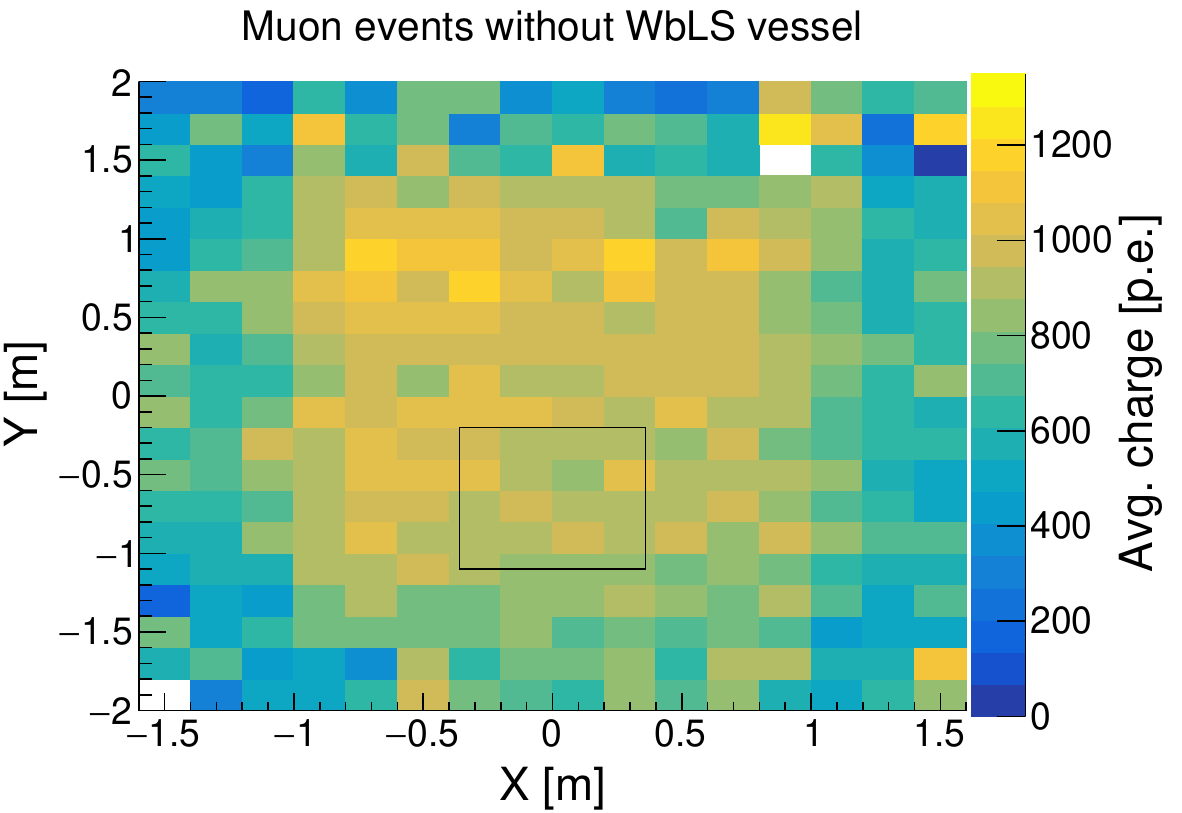}
        \caption{}
        \label{fig:muon_pos_vs_charge_presandi}
    \end{subfigure}
    \caption{The average number of detected upstream photoelectrons plotted against the intersection of the reconstructed muon track with the X-Y-plane for $\text{Z}=0$. The black outline shows the position and dimensions of the \wbls vessel. (a) For throughgoing muon events of ANNIE with the \wbls vessel. (b) For throughgoing muon events of ANNIE without the \wbls vessel.}
    \label{fig:muon_pos_vs_charge}
\end{figure}
A supplementary investigation for the principal detection of the \wbls scintillation is shown in Figure~\ref{fig:muon_pos_vs_charge}.
Here, all throughgoing muons are selected, not only those which have a reconstructed track that intersects the \wbls volume.
The plots shows the number of detected upstream photoelectrons, averaged for each position bin and plotted against the intersection of the reconstructed muon track with the X-Y-plane for the center of the water tank ($\text{Z}=0$).
%Regions with $-1.2\,\mathrm{m} < $ Y $ < 1.4\,\mathrm{m}$\,m have a low MRD efficiency and large statistical uncertainties.
The \wbls volume, centered at $\text{X}=0$\,m and $\text{Y}=-0.65$\,m, is shown as a black frame.
For the SANDI data in Figure~\ref{fig:muon_pos_vs_charge_sandi} the events with the highest average number of detected upstream photoelectrons correspond largely to the position of the \wbls-filled SANDI vessel, smeared by the resolution of the track reconstruction.
The comparison with the pre-SANDI data in Figure~\ref{fig:muon_pos_vs_charge_presandi} shows the absence of this feature for the same volume without the \wbls.
This shows that the \wbls-filled SANDI vessel is responsible for the increase in the detected number of upstream photoelectrons, which meets the expectations of the isotropic \wbls scintillation.
The structure around $X\approx -0.3$\,m, $Y\approx -1.5$\,m in Figure~\ref{fig:muon_pos_vs_charge_sandi} is a combination of a known low MRD efficiency and a low event statistic for regions with Y $ < -1.2\,\mathrm{m} < $, Y $ > 1.4\,\mathrm{m}$.
%The structure of the relatively large average charge around $X\approx -0.3$\,m, $Y\approx -1.5$\,m in Figure~\ref{fig:muon_pos_vs_charge_sandi} is a combination of known MRD efficiency effects and the low event statistics per bin in this region, with a relative statistical uncertainties of up to $\sim25\%$.

\begin{figure}[htb!]
    \centering
    \begin{subfigure}[b]{0.49\textwidth}
        \centering
        \includegraphics[width=\textwidth]{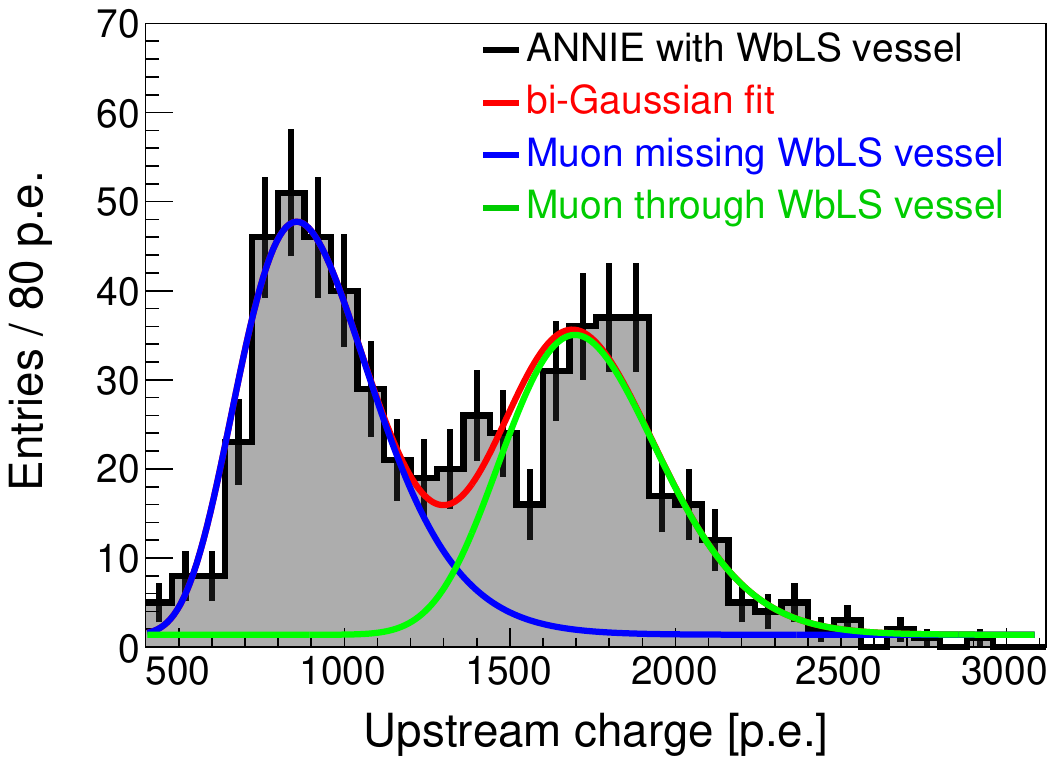}
        \caption{}
        \label{fig:comparison_up_fit}
    \end{subfigure}
    \begin{subfigure}[b]{0.49\textwidth}
        \centering
        \includegraphics[width=\textwidth]{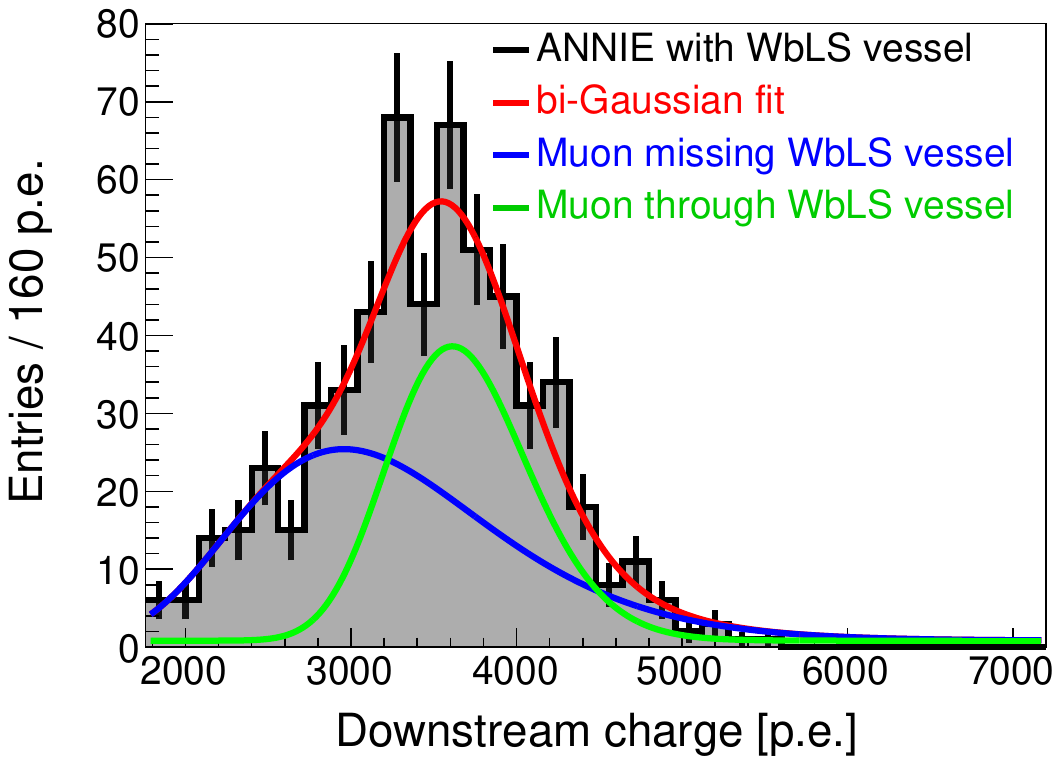}
        \caption{}
        \label{fig:comparison_down_fit}
    \end{subfigure}
    \caption{Illustration of the bi-Gaussian charge fit. The data with the \wbls vessel is shown in black, the full fit is shown in red and the two event populations of muons passing through the \wbls vessel or missing it are shown in green and blue, respectively.}
    \label{fig:throughgoing_fit}
\end{figure}

Figure~\ref{fig:throughgoing_comparison} makes it possible to provide a rough estimate of the increased light output due to the \wbls, using an analytical fit of the p.e.~distributions.
For the SANDI data, two bi-Gaussian fits are performed on the upstream and downstream p.e.~distributions. The same events that have a clear scintillation contribution in the upstream p.e.~distribution must also provide scintillation to the downstream p.e.
Therefore, the upstream p.e.~distribution can be used to estimate the number of events with and without a scintillation contribution, which can then be fixed for the fit of the downstream p.e.~distribution.
The bi-Gaussian fit on the number of photoelectrons $Q$ is defined as follows:

\begin{equation}
    f(Q) = C + \frac{A_\text{W}}{\sqrt{2\pi Q}\sigma_\text{W}} \exp{\left( -\frac{1}{2}\left(\frac{Q-\mu_\text{W}}{\sigma_\text{W} \sqrt{Q}} \right)^2\right)} + \frac{A_\text{S}}{\sqrt{2\pi Q}\sigma_\text{S}} \exp{\left( -\frac{1}{2}\left(\frac{Q-\mu_\text{S}}{\sigma_\text{S} \sqrt{Q}}\right)^2 \right)}
    \label{eq:bi_gaussian_muons}
\end{equation}
$C$ is a constant offset, $A_\text{W}$, $A_\text{S}$ are the amplitudes, $\mu_\text{W}$, $\mu_\text{S}$ are the mean values, and $\sigma_\text{W}$, $\sigma_\text{S}$ are the standard deviations for the two different event populations.
\textit{S} denotes those events that go through the \wbls vessel and produce scintillation and \textit{W} denotes those events that do not.
The additional term of $\sqrt{Q}$ takes into account the impact on the p.e.~distribution smearing due to the Poissonian nature of the number of photoelectrons.
The fit parameters of interest are the mean p.e. values of the two event populations: The number of upstream p.e. $\mu_\text{W}^\text{U} = (880 \pm 20)$\,p.e. and the number of downstream p.e. $\mu_\text{W}^\text{D} = (3056 \pm 59)$\,p.e. for those muon events that do not cross SANDI and the corresponding values $\mu_\text{S}^\text{U} = (1713 \pm 34)$\,p.e., $\mu_\text{S}^\text{D} = (3637 \pm 51)$\,p.e. for those muon events that do cross SANDI and produce scintillation light.
The best fit results are also illustrated in Figure~\ref{fig:throughgoing_fit} for the SANDI data.
The black histogram shows the data, the red line shows the best fit distribution, and the blue and green lines show the p.e. distributions of the fit corresponding to the two event populations for illustration purposes.
The pre-SANDI data is fitted with a single Gaussian, resulting in $\mu_\text{pre-SANDI}^\text{U} = (887 \pm 9)$\,p.e., $\mu_\text{pre-SANDI}^\text{D} = (3473 \pm 25)$\,p.e.

The muons that cross the \wbls vessel are not fully contained inside it and the above fit values include a large fraction of Cherenkov light from pure water as well as scintillation and Cherenkov light from the \wbls vessel.
Additionally, the \wbls can absorb Cherenkov light, without re-emitting scintillation, and the SANDI structure can also have a shadowing effect.
It is possible to use the above p.e. mean values and the MRD reconstructed muon tracks to correct for the muon tracklength through the \wbls vessel and estimate the loss of Cherenkov light due to SANDI.
For this purpose a simple toy-MC simulation has been used to estimate the average true track length $L$ of the muons, given the reconstructed MRD tracks and their uncertainties. 
The estimated value for the average track length through the entire ANNIE water tank is $l_\text{ANNIE} = (2.951 \pm 0.012)\,\text{m}$.
A small selection effect has been observed on this value for the two event populations, where the muon tracks missing the \wbls vessel give a value of $l_\text{W} = (2.896 \pm 0.012)\,\text{m}$ and the events crossing it give $l_\text{S} = (3.019 \pm 0.011)\,\text{m}$.
The reason for this effect is the cylindrical shape of the water tank, where muon tracks that miss the \wbls vessel tend to intersect the water tank more often at off-center positions.
The average track length of the muons through the \wbls vessel is $l_\text{\wbls} = (0.597 \pm 0.007)\,\text{m}$.
The toy-MC has also been used to estimate the average track length of the Cherenkov photons through the \wbls vessel, which are emitted along the muon tracks, as $l_\text{che} = (0.463\pm0.010)\,\text{m}$.
The average fraction of the Cherenkov photons that pass through the \wbls vessel is estimated likewise as $r_\text{che} = 0.123\pm0.003$.

In the following we define $q_\text{W}$ [p.e.\,/\,m] as the estimated number of detected Cherenkov photoelectrons per meter of muon track in pure water, $q_\text{S}$ [p.e.\,/\,m] as the estimated number of Cherenkov and scintillation photoelectrons from the \wbls, and $A_\text{eff}$ as the effective absorption of the Cherenkov light by the SANDI vessel, its steel holding structure and the \wbls itself.
The following equations can be used to estimate these parameters, assuming that the total number of detected photoelectrons is simply proportional to the track length of the muon through the medium.  The relevant proportionality constants are $q_\text{W}$ and $q_\text{S}$:
\begin{equation}
\begin{split}
    \mu_\text{pre-SANDI}^\text{U} + \mu_\text{pre-SANDI}^\text{D} &= q_\text{W} \cdot l_\text{ANNIE}\\
    \mu_\text{W}^\text{U} + \mu_\text{W}^\text{D} &= q_\text{W} \cdot l_\text{W}\cdot \left(1 - A_\text{eff}\right) \\
    \mu_\text{S}^\text{U} + \mu_\text{S}^\text{D} &= q_\text{W} \cdot l_\text{S}\cdot \left(1 - A_\text{eff}\right) + q_\text{S}\cdot l_\text{\wbls}
\end{split}
\end{equation}
Solving the above equations results in $q_\text{W} = (1395 \pm 10)\,\text{p.e./m}$ and $q_\text{S} = (2159 \pm 154)\,\text{p.e./m}$, where the errors have been added in quadrature.
Therefore, the ratio for the estimated number of photoelectrons per meter of muon track length from the \wbls compared to the number of photoelectron from pure water is $\frac{q_\text{S}}{q_\text{W}} = 1.42 \pm 0.13$.

Additionally, the equations allow for an estimation of the total absorption of Cherenkov light. The absorption parameter $A_\text{eff}$ is an effective parameter, which includes all possible sources of the Cherenkov absorption due to the deployment of SANDI, such as the acrylic SANDI vessel, its steel holding structure and the \wbls attenuation of the Cherenkov light.
This value is estimated as $A_\text{eff} = 0.08 \pm 0.03$, meaning that roughly 8\% of the Cherenkov light is missing due to the \wbls-filled SANDI vessel, compared to the normal configuration of the ANNIE water tank.

A number of potential sources of systematic uncertainty have been investigated for the above calculations.
First, the position of SANDI inside the ANNIE water tank is known only with a certain precision. The SANDI position has been estimated through the bi-Gaussian fit of Eq.~\ref{eq:bi_gaussian_muons} and the comparison of the relative contribution of true scintillation events $A_\text{S}/A_\text{W}$ for different assumed positions of SANDI.
The corresponding SANDI position uncertainty has been estimated conservatively as $\pm0.1\,\text{m}$.
Re-performing the calculations again for different assumed SANDI positions results in a systematic uncertainty of $\Delta \frac{q_\text{S}}{q_\text{W}} = \pm0.16$.
Second, the analytical fit has also been re-performed with some variations, such as leaving out the sliding standard deviation term $\sqrt{Q}$, without the constant $C$, and within different fit ranges. The corresponding systematic uncertainty is estimated as $\Delta \frac{q_\text{S}}{q_\text{W}} = \pm0.07$.
Adding these uncertainties in quadrature results in a ratio of $\frac{q_\text{S}}{q_\text{W}} = 1.42 \pm 0.13\text{\,(stat.)} \pm 0.18\text{\,(syst.)} = 1.42 \pm 0.23\text{\,(stat. + syst.)}$.

The ratio of the muon tracks that go through SANDI relative to all selected events can also be estimated using the fit.
The result is $0.44\pm0.13\text{\,(stat. + syst.)}$, where the systematic uncertainty is estimated in the same as described above.
This corresponds to about 270 muons passing through the SANDI vessel.

Last, it has to be noted here that the ratio $\frac{q_\text{S}}{q_\text{W}}$ is a simple measurement of the expected number of photoelectrons with and without the \wbls-filled SANDI vessel, given the selected throughgoing muons events.
This is not a measurement of the intrinsic \wbls scintillation light-yield.
Such an in-situ measurement necessitates a dedicated, full Monte Carlo based analysis, which takes into account the full detector response and geometry, as well as all interactions of the photons with the different detector media.

\subsection{Neutrino candidate events}
\begin{figure}[htb!]
    \centering
    \begin{subfigure}[b]{0.49\textwidth}
        \centering
        \includegraphics[width=\textwidth]{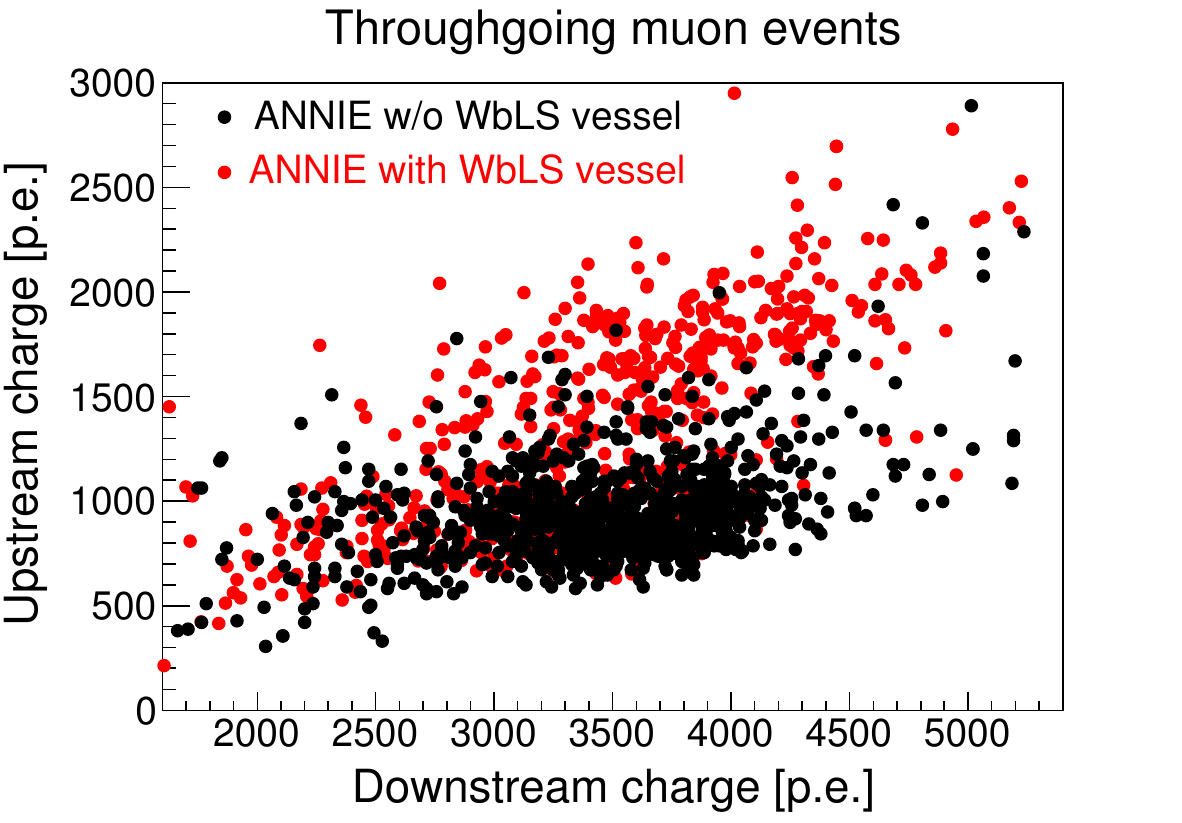}
        \caption{}
        \label{fig:throughgoing_down_vs_up_scatterplot}
    \end{subfigure}
    \hfill
    \begin{subfigure}[b]{0.49\textwidth}
        \centering
        \includegraphics[width=\textwidth]{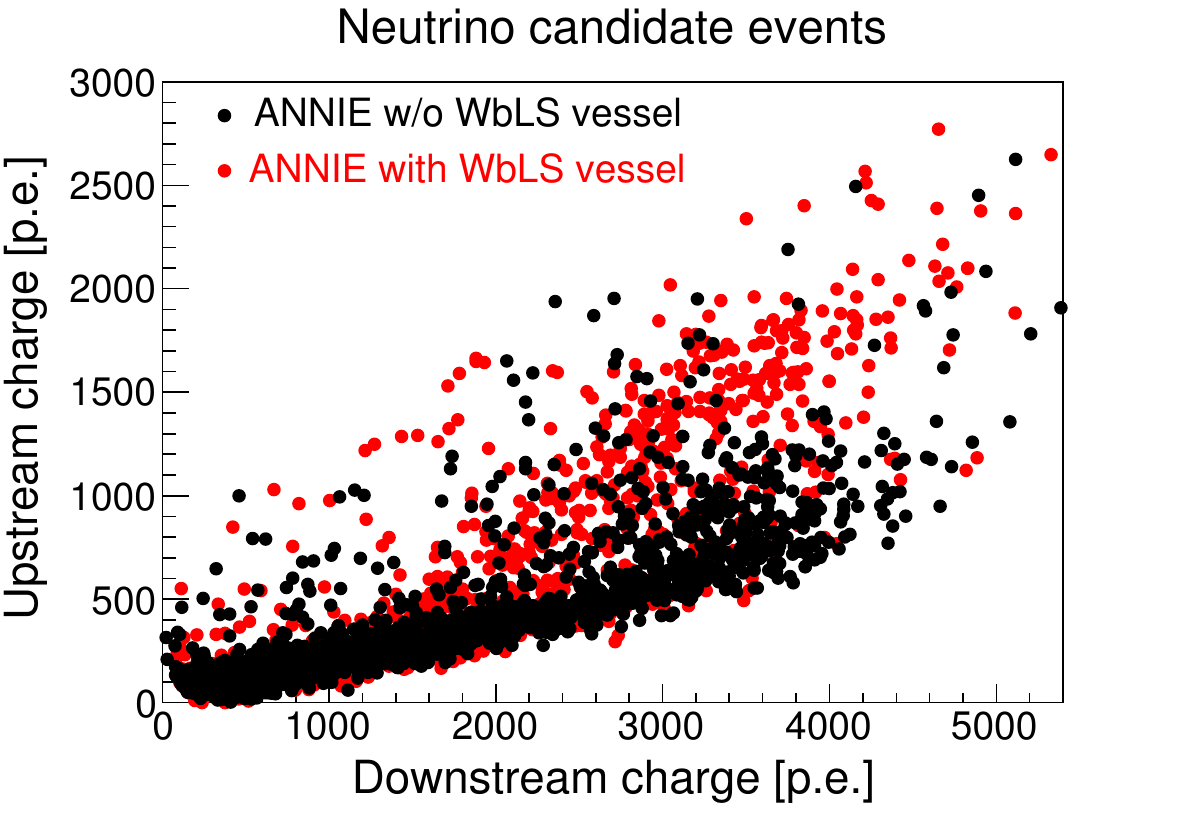}
        \caption{}
        \label{fig:beamneutrinos_down_vs_up_scatterplot}
    \end{subfigure}
    \caption{Scatter plots of the number of detected photoelectrons for the downstream PMTs vs. the upstream PMTs for the selected events of ANNIE without the \wbls vessel in black and for the events of ANNIE with the \wbls vessel in red. (a) Throughgoing muon events, which required a coincident FMV trigger. (b) Neutrino candidate events, which required the absence of a coincident FMV trigger.}
    \label{fig:scatterplots_comparison}
\end{figure}

The next part of this section shows the \wbls scintillation light for neutrino candidate events in ANNIE.
These beam neutrino events have been selected in the same way as the throughgoing muon events, with the difference that now the FMV is required to have not seen a signal in coincidence with the tank PMTs and the MRD.
This selects muon events that have been produced through neutrino interactions inside the ANNIE tank.
The MRD reconstructed muon track is again required to intersect the \wbls vessel.

Figure~\ref{fig:scatterplots_comparison} shows scatter plots of the number of detected photoelectrons for the downstream PMTs against the upstream PMTs, where the pre-SANDI data without the \wbls vessel is shown in black, while SANDI data with the \wbls vessel is shown in red.
Throughgoing muon events are shown in Figure~\ref{fig:throughgoing_down_vs_up_scatterplot}.
The same features described in Figure~\ref{fig:throughgoing_comparison} are also visible here.
There are again two populations of events for the SANDI data, where the mis-reconstructed muon track events without scintillation are hidden in the plot behind the black points.
The events with scintillation show a stronger linear dependency between the upstream and downstream photoelectrons compared to the Cherenkov light from ANNIE without \wbls.
This is an expected behavior from the isotropy of scintillation.

Figure~\ref{fig:beamneutrinos_down_vs_up_scatterplot} shows the neutrino candidate events, i.e. neutrino interactions which produce muons inside the water tank, as selected by the FMV.
For a downstream photoelectron number below $1500$\,p.e. there is no clear difference between the pre-SANDI and SANDI data events. 
As the number of downstream photoelectrons is proportional to the muon path length through the water tank this likely means that these muons are produced in neutrino interactions behind the \wbls vessel.
For larger numbers of downstream photoelectrons the SANDI data shows again two event populations, as described previously.
The events which produce scintillation light have a distribution which is well distinguishable from the events of ANNIE without the \wbls.
This scatter plot therefore provides a qualitative proof that \wbls produces a larger amount of light than pure water for accelerator neutrino events in an operating large scale neutrino detector.

\subsection{Michel electron analysis}
Michel electrons, produced by decaying muons, follow a well known spectrum that can be used as a standard candle for the comparison of the detected number of photoelectrons in water and in the \wbls.
The total energy spectrum of Michel electrons in vacuum can be described as follows~\cite{michel_parameter}:

\begin{equation}
    N(E) = 3\left(\frac{E}{E_\text{max}}\right)^2 - 2\left(\frac{E}{E_\text{max}}\right)^3 \label{eq:c_spectrum},
\end{equation}
where E is the energy of electron and $E_\text{max}\approx 53\,$MeV is the maximal possible electron energy in the decay. 

%The contribution of anti-muons from the decay of anti-muon neutrinos is considered negligible for the following selection of Michel electrons.
A coincident event pair is required for the Michel electron selection: A prompt event caused by a muon, followed by a second event caused by the Michel electron. The prompt event is selected by several cuts to exclude most of the background: First, the prompt event is required to be triggered within the 2\,${\mu}$s acquisition window and it is required to be the brightest signal among all the signals within a time window of the corresponding BNB trigger. Second, a FMV hit is required to be in coincidence with the muon candidate event to enforce that the muon source is located outside of the ANNIE detector. Furthermore, a prompt event is required to have a barycenter located in the downstream part of the detector and the event time is required to have no MRD signal within a 50\,ns time window.
This cut enforces the selection of muons stopping in the tank.
Finally, a photoelectron range of [1000, 4000]\,p.e. is required to further exclude muon candidates that are falsely tagged as stopping muons to compensate for the MRD inefficiency. In the case that an event passes all the cuts mentioned above, all subsequent clusters within a time window of ${\Delta}t$ $\in$ [1000, 6000]\,${\mu}$s are tagged as Michel electron candidate events. A charge balance cut of $q_\text{cb}$ < 0.2 is applied to the Michel electron candidate events to further reduce background caused by a large, single PMT signal. The charge balance parameter $q_\text{cb}$ is defined using the following equation: 
\begin{equation}
q_\text{cb} = \sqrt{\frac{{\Sigma}Q^2_i}{Q_\text{sum}^2}-\frac{1}{N}} ,
\end{equation}
where $Q_\text{sum}$ represents the total number of observed photoelectrons in the cluster, $Q_i$ references the number of photoelectrons that have been observed by $\text{PMT}_i$, and $N$ is the number of PMTs involved in the cluster. Charge balance values close to 1 indicate that most of the charge has been seen by a single PMT, indicating that the cluster was probably a noise event.

\begin{figure}[htb!]
    \centering
    \begin{subfigure}[b]{0.49\textwidth}
        \centering
        \includegraphics[width=\textwidth]{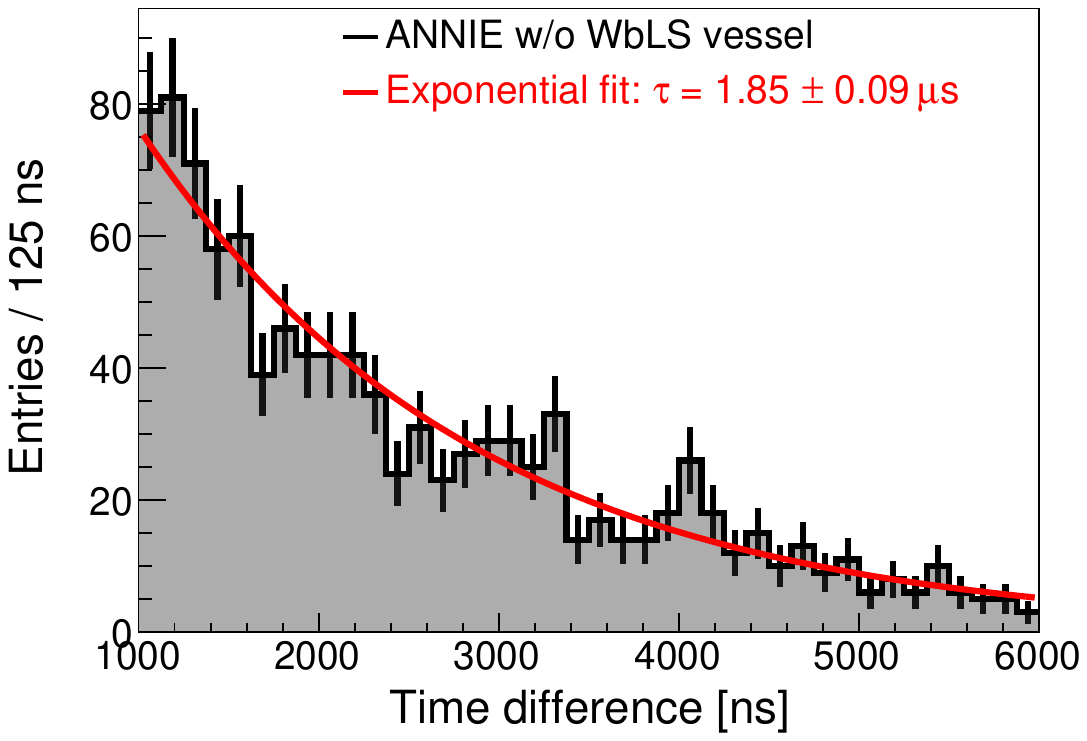}
        \caption{}
        \label{fig:presandi_michel_time}
    \end{subfigure}
    \begin{subfigure}[b]{0.49\textwidth}
        \centering
        \includegraphics[width=\textwidth]{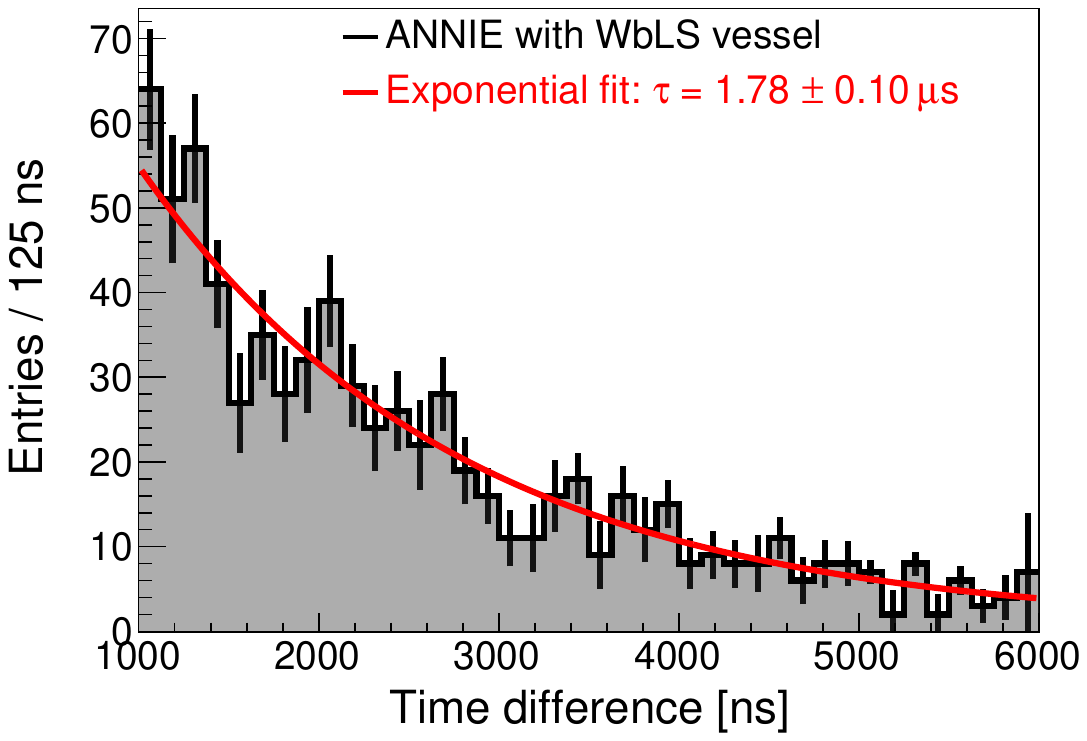}
        \caption{}
        \label{fig:sandi_michel_time}
    \end{subfigure}
    \caption{Histograms for the timing distribution between Michel electron events and the preceding muon events. They follow the exponential decay rule for (a) the data without the \wbls vessel and (b) the data with the \wbls vessel.}
    \label{fig:michel_time}
\end{figure}
A good comparison of the number of photoelectrons produced by Michel electrons in pure water and in the SANDI vessel requires a pure sample with minimum background.
As a measure of sample purity, the time distribution of the selected Michel electron events is shown in Figure~\ref{fig:michel_time}. 
The pre-SANDI data is shown on the left and the SANDI data is shown on the right.
The timing spectrum is fitted by a function of an exponential decay plus noise that is assumed to be a constant:
\begin{equation}
    A = A_0 e^{-t/\tau} + C ,
\end{equation}
where $A_0$ denotes the initial amplitude at $t=0$, $\tau$ is the decay constant and $C$ denotes the noise. 
The fitted decay constants are $\tau = (1.78\pm0.10)\,{\mu}$s for the SANDI data and $\tau = (1.85\pm0.09)\,{\mu}$s for the pre-SANDI data.
The expected muon lifetime is calculated using the following equation:
\begin{align}
            \frac{1}{\tau_\text{tot}} = R_\text{tot} &= R_\text{vac} + R_\text{oxy},
\end{align}
where ${R_\text{tot}}$ denotes the total muon decay rate, ${R_\text{vac}}$ and ${R_\text{oxy}}$ denote the muon decay rate in vacuum and the capture rate in pure oxygen, respectively. $\tau_\text{tot}$ is the overall lifetime of the muon.
From the experimental results of \cite{mu_capture_on_nuclei}, the expected muon lifetime $\tau_\text{tot}$ is $(1.788\pm0.002){\mu}$s, which is well in agreement with the results of both event selections.
The effect of muon capture on hydrogen is ignored here. 

\begin{figure}[htb!]
    \centering
    \begin{subfigure}[b]{0.49\textwidth}
        \centering
        \includegraphics[width=\textwidth]{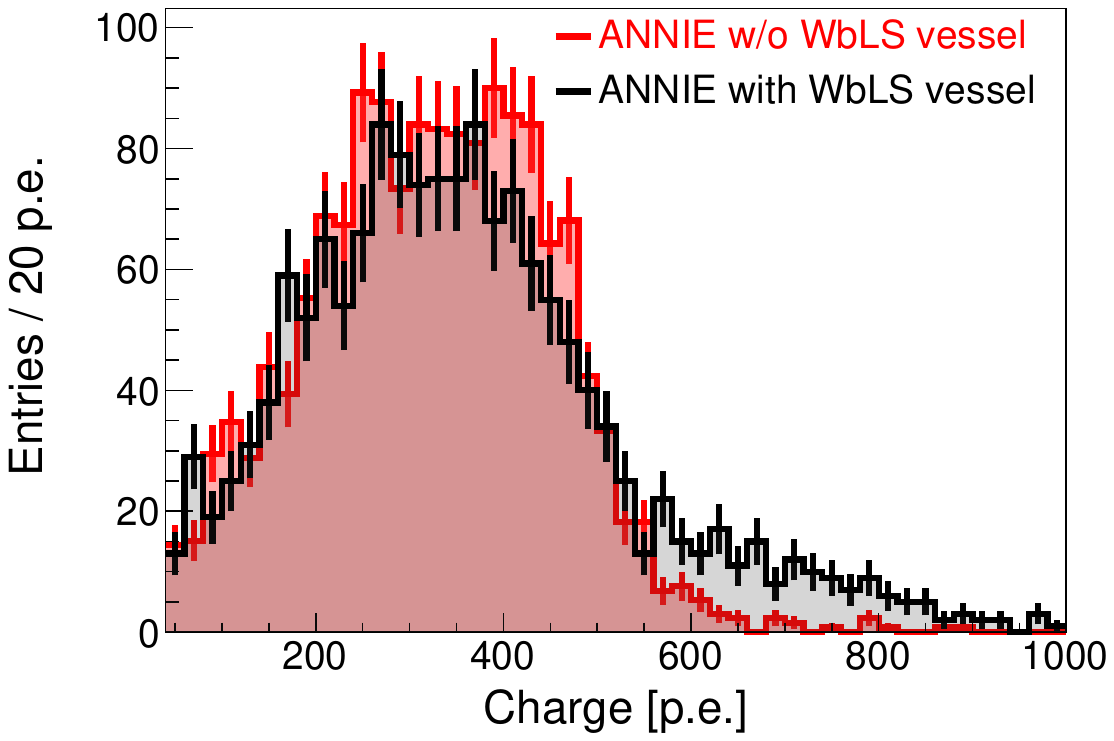}
        \caption{}
        \label{fig:michel_charge_compare}
    \end{subfigure}
    \begin{subfigure}[b]{0.49\textwidth}
        \centering
        \includegraphics[width=\textwidth]{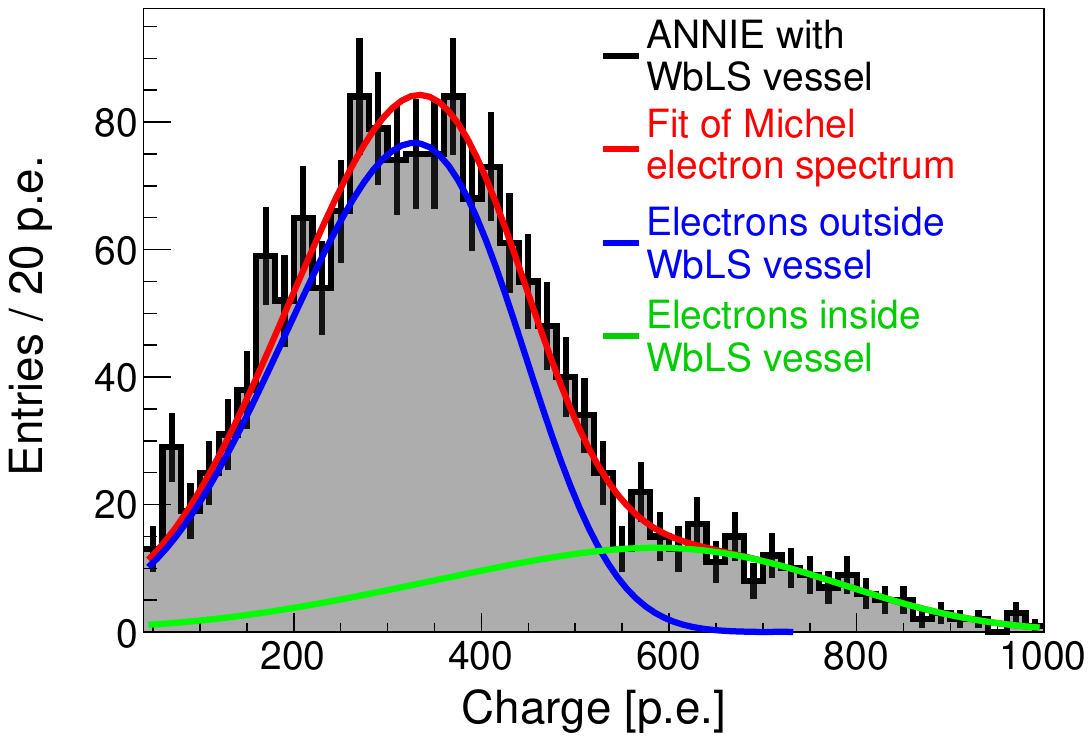}
        \caption{}
        \label{fig:sandi_michel_charge}
    \end{subfigure}
    \caption{Histograms of the Michel electron charge distribution collected by PMTs inside the ANNIE tank. (a) The data of ANNIE without the \wbls vessel in red and for ANNIE with the \wbls in black. The ANNIE without \wbls data is normalized to the statistics of ANNIE with the \wbls vessel for better comparison. (b) The fit of the Michel electron data for ANNIE with the \wbls vessel. This is a combination of events inside and outside of the \wbls vessel, estimated by the fit and shown in blue and green, respectively.}
    \label{fig:michel_charge}
\end{figure}
Figure~\ref{fig:michel_charge_compare} shows the charge distribution of the selected Michel electron events for the pre-SANDI data in red and for the SANDI data in black.
It can be seen, that the selected Michel electrons fall into two categories: Those electron events for which the parent muon decayed within the SANDI vessel and those electron samples for which the parent muon decayed outside of the SANDI vessel. 
For those Michel electrons that have been generated outside of SANDI the source of the detected photoelectrons is from Cherenkov radiation, while for the ones generated inside SANDI the source is a combination of both Cherenkov radiation and \wbls scintillation.
The latter event category can be seen in comparison with the pre-SANDI Michel electron charge distribution above $\sim600\,$p.e.

The comparison of the mean values of these two event populations of the SANDI Michel electrons can provide an estimation of the additional number of photoelectrons provided by the deployment of the \wbls-filled vessel.
A Gaussian smeared Michel spectrum is used to describe the total number of photoelectrons  collected by the PMTs inside ANNIE tank $Q$. The spectrum is described by the following equation:

\begin{equation}
    g(Q) = \int_{0}^{E_m} A\cdot f(E)\cdot G\left(Q,\mu(E),\sigma\right) dE,
    \label{eq:smeared_spectrum}
\end{equation}
where $f(E){\propto}N(E)$ is defined in equation \eqref{eq:c_spectrum} as the Michel electron energy spectrum, $G(Q,\mu(E),\sigma)$ is the Gaussian distribution, $\mu(E)=k E$ is the mean number of photoelectrons as a function of energy and is naively assumed to be proportional to the energy deposited inside the ANNIE tank. $\sigma$ is the energy resolution of the detector and is a constant here. 

Equation \eqref{eq:smeared_spectrum} is then used to fit the photoelectron distribution in Figure~\ref{fig:sandi_michel_charge}.
Here the charge distribution of the SANDI data is a combination of the two event populations described above, each with a unique mean value for the number of detected photoelectrons.
The parameter of interest is the proportionality factor $k$ of the photoelectron mean value for the two distributions: $k_\text{SANDI} = (14.44 \pm 0.87)$\,p.e./MeV and $k_\text{Water} = (8.17 \pm 0.14)$\,p.e./MeV.
The correlation coefficient of these parameters given by the fit is $0.0073$.
The ratio of $k_\text{SANDI} / k_\text{Water} = 1.77 \pm 0.06$  can be used to estimate the relative increase of the detected number of photoelectrons in ANNIE due to the \wbls-filled SANDI vessel, compared to pure water.
Like with the previous analysis of throughgoing muons the systematic uncertainty here has been estimated by varying the analytical fit function; with and without a sliding standard deviation $\sigma = \sigma_0\sqrt{E}$. 
The corresponding uncertainty is $\Delta \frac{k_\text{SANDI}}{k_\text{Water}} = \pm0.05$, resulting in $\frac{k_\text{SANDI}}{k_\text{Water}} = 1.77\pm0.06\text{\,(stat.)}\pm0.05\text{\,(syst.)} = 1.77\pm0.08\text{\,(stat.+syst.)}$.

\section{Conclusions}
\label{sec:conclusion}
\wbls is a novel target material which allows for the simultaneous detection of Cherenkov radiation and scintillation.
Cherenkov radiation provides sensitivity to the particle direction and track reconstruction, as well as allowing for particle identification through the Cherenkov ring topology. The addition of scintillation improves the energy resolution and allows for the detection of particles below the Cherenkov threshold.
The work described in the present paper is part of the ongoing technical demonstration and experimental validation of the hybrid detection concept, which is important for future, large scale, advanced optical neutrino detectors such as THEIA~\cite{theia_2020}.

In this paper we present the first implementation of \wbls in a running neutrino beam, using a system dubbed SANDI (Scintillator for ANNIE Neutrino Detection Improvement).
SANDI, a cylindrical vessel with a diameter of $72$\,cm and $90$\,cm height, was successfully filled with \wbls and deployed in the ANNIE water tank for about two months.
Samples of the \wbls have been measured at Brookhaven National Laboratory before and after the SANDI deployment.
No deterioration of optical transparency or light-yield have been found, which indicates a long term stability of the \wbls under real deployment and beam measurement conditions.

The \wbls scintillation signal can be quantified by the comparison of the ANNIE data with the \wbls vessel and of ANNIE data without the SANDI vessel.
Scintillation has been observed for throughgoing muons, Michel electrons from muon decays inside ANNIE, as well as beam neutrino interactions inside ANNIE.
Two independent estimations of the increase of the detected number of photoelectrons due to the deployment of the \wbls vessel have been performed, one with throughgoing muons and one with Michel electron events.
These analyses have been performed through analytical fits on the distributions of the detected number of photoelectrons.
The analyses results are expressed as a ratio of the number of detected photoelectrons for events that deposit energy within the \wbls-filled SANDI vessel and the detected number of photoelectrons for events outside of the SANDI vessel.
These ratios are $q_\text{SANDI}/q_\text{Water}= 1.42\pm0.23\text{\,(stat.+syst.)}$ for the throughgoing muons and $k_\text{SANDI}/k_\text{Water} = 1.77\pm0.08\text{\,(stat.+syst.)}$ for the Michel electrons.

The observed increase in the detected number of photoelectrons is clearly caused by the additional scintillation of the \wbls and the order of magnitude of this detected increase is in agreement with the expectations given by bench-top experiments~\cite{chess_2023_prep}.
Nonetheless, it has to be noted that above results cannot be directly interpreted as a measurement of the intrinsic scintillation light-yield of the \wbls. 
They rather express the increase in the number of detected photoelectrons with and without the deployment of the \wbls-filled SANDI vessel.
An in-situ measurement of the intrinsic \wbls light-yield requires a detailed Monte Carlo model of the full ANNIE detector response that includes the effects of the SANDI vessel structure and the absorption, re-emission and scattering of light in the \wbls.
The work of producing such a Monte Carlo model is currently underway.

The results presented here will form the basis for the further development of hybrid event detection in the GeV energy range in ANNIE. Future reconstruction and particle identification algorithms will make use of both the Cherenkov light and the \wbls scintillation.
These algorithms are expected to depend on critical parameters of the \wbls, such as its attenuation length and scintillation light-yield.
Using the upgraded reconstruction techniques, the planned follow-up analyses on the collected \wbls data regard the detection of neutron captures on hydrogen, the search for neutral current events, as well as the investigation of the hadronic scintillation component, all within the \wbls vessel. Further plans include the re-deployment of SANDI with gadolinium-loaded \wbls to study the enhanced neutron detection and potentially filling the entire ANNIE tank with \wbls to permit a better assessment of the performance of hybrid event reconstruction in a future large-scale experiment.

\acknowledgments

This work was supported by: (1) the U.S. Department of Energy (DOE) Office of Science under award number DE-SC0009999 (UC Davis) and award number DE-SC0015684 (Iowa State University), and by the DOE National Nuclear Security Administration through the Nuclear Science and Security Consortium under award number DE-NA0003180 (UC Davis); (2) Deutsche Forschungsgemeinschaft grants 456139317 (Hamburg) and 490717455 (Mainz and T\"ubingen); (3) National Science Foundation Grant No. PHY-2310018 (Ohio State), PHY-2047665 (Rutgers), and OIA-2132223 (South Dakota Mines); (4) UK Research and Innovation FLF MR/S032843/1 (Warwick); (5) Scientific Research Projects (BAP) of Erciyes University under the grant numbers of of FBAU-2023-12325, FDS-2021-11525 and FBA-2022-12207.
Work conducted at Lawrence Berkeley National Laboratory was performed under the auspices of the U.S. Department of Energy under Contract DE-AC02-05CH11231.
Work conducted at Lawrence Livermore National Laboratory was performed under the auspices of the U.S. Department of Energy by Lawrence Livermore National Laboratory under contract DE-AC52-07NA27344, release number LLNL-JRNL-858196.
The work conducted at Brookhaven National Laboratory was supported by the U.S. Department of Energy under contract DE-AC02-98CH10886. 
The project was funded by the U.S. Department of Energy, National Nuclear Security Administration, Office of Defense Nuclear Nonproliferation Research and Development (DNN R\&D). 
Finally, we gratefully acknowledge all the Fermilab scientists and staff who supported this work through their technical expertise and operational assistance at the Booster Neutrino Beam.

\bibliographystyle{JHEP}
\bibliography{sandi_paper.bib}

\end{document}